\documentclass[aps,preprint,nofootinbib,superscriptaddress]{revtex4}

\usepackage{mathrsfs}
\usepackage{amsmath}
\usepackage{array}
\usepackage{verbatim}
\usepackage{epsfig}
\usepackage{epsf}
\usepackage{CJK}
\usepackage{graphicx}
\begin{document}

\title{Dynamical nucleus-nucleus potential at short distances}

\author{Yongying Jiang  }
\affiliation{Department of Physics, Guangxi Normal University,
Guilin 541004, P. R. China}

\author{Ning Wang \footnote{Corresponding author \\ {\it E-mail address}: wangning@gxnu.edu.cn}}
\affiliation{Department of Physics, Guangxi Normal University,
Guilin 541004, P. R. China}

\author{Zhuxia Li  }
\affiliation{China Institute of Atomic Energy, Beijing 102413, P.
R. China}

\author{Werner Scheid}
\affiliation{Institute for Theoretical Physics at
Justus-Liebig-University, D-35392 Giessen, Germany}

\begin{abstract}
The dynamical nucleus-nucleus potentials for fusion reactions
$^{40}$Ca+$^{40}$Ca, $^{48}$Ca+$^{208}$Pb and
$^{126}$Sn+$^{130}$Te are studied with the improved quantum
molecular dynamics (ImQMD) model together with the extended
Thomas-Fermi approximation for the kinetic energies of nuclei. The
obtained fusion barrier for $^{40}$Ca+$^{40}$Ca is in good
agreement with the extracted fusion barrier from the measured
fusion excitation function, and the depth of the fusion pockets
are close to the results of time-dependent Hartree-Fock
calculations. The energy dependence of fusion barrier is also
investigated. For heavy fusion system, the fusion pocket becomes
shallow and almost disappears for symmetric systems and the
obtained potential at short distances is higher than the adiabatic
potential.

%{\bf PACS numbers:}~25.70.-z, 24.10.-i\\
%{\bf Key Words:}~ImQMD model, ETF approximation, dynamical fusion
%barrier, nucleus-nucleus potential }\\
\end{abstract}
\maketitle

\begin{center}
\textbf{I. INTRODUCTION}
\end{center}

The synthesis of super-heavy elements (SHEs) has been studied for
many years both theoretically and experimentally
\cite{Ada98,Shen,SHG00,YTO04}. Up to now the super-heavy nuclei
are uniquely synthesized through fusion reaction including "cold"
fusion reaction with lead and bismuth targets \cite{SHG00} and
"hot" fusion with actinide targets \cite{YTO04}. The study of the
dynamical process in fusion reactions especially the
nucleus-nucleus potential is of great importance for the synthesis
of SHEs. Experimentally, the fusion barrier distributions can be
directly obtained from the measured fusion excitation functions,
with which the information of the nucleus-nucleus potential around
the fusion barrier can be obtained. Fig.1(a) shows the
nucleus-nucleus potential calculated with different models for
$^{40}$Ca+$^{40}$Ca. We can see that the obtained barrier heights
with different models are close to each other and all of them are
comparable with the extracted mean barrier height, while the
calculated nucleus-nucleus potentials at short distances are quite
different with different models. It is known that the adiabatic
and diabatic approximations lead to different nucleus-nucleus
potentials especially at short distances and thus to different
fusion paths and different mechanisms of fusion reactions. Both
approximations are frequently applied to the study of the
synthesis of super-heavy nuclei \cite{Ada98,Zag}. For
understanding the fusion mechanism of a heavy system, it is
important and necessary to study the nucleus-nucleus potential at
short distances, with which one could get information on the
fusion path and the formation probability of the di-nuclear system
in reactions leading to super-heavy nuclei.

For description of heavy-ion fusion reactions, some theoretical
models have been developed. The fusion coupled channel model is a
powerful tool to calculate the fusion excitation function and to
investigate the influence of nuclear structure effects on the
fusion cross sections \cite{Hag99,Ich07,Esb07}. On the other hand,
the microscopic dynamics model such as time dependent Hartree Fock
(TDHF) model \cite{Umar06, KW08} and the improved quantum
molecular dynamics (ImQMD) model \cite{WN02,WN04} are widely
applied to study the dynamical behavior of fusion process. The
ImQMD model is a semi-classical microscopic dynamics model and is
successfully used for intermediate-energy heavy-ion collisions and
for heavy-ion collisions at energies near the Coulomb barrier
\cite{WN02,WN04,WN05,Zhang08}. In the ImQMD model the dynamical
effects such as the dynamical deformation, neck formation, etc,
are microscopically and self-consistently taken into account.
Recently, the model has been applied to the study of the dynamical
barrier of a heavy system \cite{TJL09}, of mass parameters
\cite{ZK09}, and of the strongly damped process of
$^{238}$U+$^{238}$U \cite{TJL08,Zhaok09}. In this paper we
carefully investigate the kinetic energies of nuclei based on the
extended Thomas-Fermi approximation with which the dynamical
fusion barrier is accurately obtained. The paper is organized as
follows: In Sec. II, the ImQMD model will be briefly introduced.
In Sec. III, some calculated results on the kinetic energies of
nuclei and the nucleus-nucleus potential for $^{40}$Ca+$^{40}$Ca,
$^{48}$Ca+$^{208}$Pb and $^{126}$Sn+$^{130}$Te will be presented.
Finally, a summary is given in Sec. IV.

\begin{figure}
\includegraphics[angle=-0,width=0.7\textwidth]{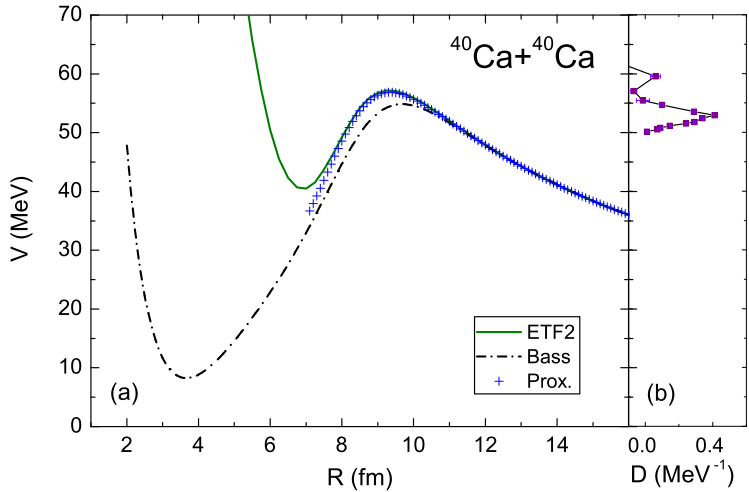}
 \caption{(Color online) (a) The nucleus-nucleus potential as a function of the center-to-center
 distance $R$ between two nuclei for the reaction $^{40}$Ca+$^{40}$Ca. The solid curve denotes the results of Skyrme energy
  density functional  together with the extended Thomas-Fermi (ETF2)
   approximation \cite{liumin06}. The dash-dotted curve and the crosses denote the Bass
potential \cite{bass80} and the proximity potential \cite{prox00},
respectively. (b) The fusion barrier distribution extracted from
the measured fusion excitation function \cite{HAA84} (see Eq.(9)
of Ref.\cite{Wang07}).}
\end{figure}

\begin{center}
\textbf{II. THE IMPROVED QUANTUM DYNAMICS MODEL}
\end{center}

In the ImQMD model, the same as in the original QMD model
\cite{Ai91}, each nucleon is represented by a coherent state of a
Gaussian wave packet. Through a Wigner transformation, one can get
the one-body phase space distribution function for
N-distinguishable particles (see \cite{Ai91,WN02} for details).
The density distribution function $\rho$  of a system reads
\begin{equation}
\rho (\mathbf{r})=\sum\limits_{i}\frac{1}{(2\pi \sigma
_{r}^{2})^{3/2}}\exp [-\frac{( \mathbf{r-r}_{i})^{2}}{2\sigma
_{r}^{2}}].
\end{equation}
Where, $\sigma _{r}$ represents the spatial spread of the wave
packet. The propagation of nucleons is governed by Hamiltonian
equations of motion under the self-consistently generated mean
field:
\begin{equation}  \label{1}
\dot{\mathbf{r}}_{i}=\frac{\partial H}{\partial \mathbf{p}_{i}}, \dot{%
\mathbf{p}}_{i}=-\frac{\partial H}{\partial \mathbf{r}_{i}}.
\end{equation}
Where, $\mathbf{r}_{i}$ and $\mathbf{p}_{i}$ are the centers of
i-th wave packet in the coordinate and momentum space,
respectively. The Hamiltonian $H$ consists of the kinetic energy
and the effective interaction potential energy:
\begin{equation}  \label{2}
H=T+U,
\end{equation}
\begin{equation}  \label{3}
T=\sum\limits_{i}\frac{\mathbf{p}_{i}^{2}}{2m}.
\end{equation}
%The contribution of the wave packet to the kinetic energy which is
%a constant ($\frac{\hbar^{2}}{2m}\frac{3N}{4\sigma_{r}^{2}}$) is
%not included in this model.
The effective interaction potential energy includes the nuclear
interaction potential energy and the Coulomb interaction potential
energy,
\begin{equation}
U=U_{loc}+U_{Coul},  \label{4}
\end{equation}%
with
\begin{equation}
U_{loc}=\int V_{loc}(\mathbf{r})d\mathbf{r}.  \label{5}
\end{equation}%
Where $V_{loc}(\mathbf{r})$ is the potential energy density which
is obtained by the effective Skyrme interaction and taken as the
same as that in Ref.\cite{WN04}
\begin{equation}
V_{loc}=\frac{\alpha }{2}\frac{\rho ^{2}}{\rho _{0}}+\frac{\beta }{\gamma +1}%
\frac{\rho ^{\gamma +1}}{\rho _{0}^{\gamma }}+\frac{g_{sur}}{2\rho _{0}}%
(\nabla \rho )^{2}+g_{\tau }\frac{\rho ^{\eta +1}}{\rho _{0}^{\eta }}+\frac{%
C_{s}}{2\rho _{0}}(\rho ^{2}-\kappa _{s}(\nabla \rho )^{2})\delta
^{2}, \label{6}
\end{equation}%
where  $\delta =\frac{\rho _{n}-\rho _{p}}{ \rho _{n}+\rho _{p}}$
is the isospin asymmetry. Inserting expression (1) together with
(7) into (6), we obtain the interaction potential energy
\begin{eqnarray}
U_{loc} &=&\frac{\alpha }{2}\sum\limits_{i} \frac{\rho _{i}}{\rho
_{0}}+\frac{\beta }{\gamma +1}\sum\limits_{i}\left(
 \frac{\rho _{i}}{\rho _{0}}\right) ^{\gamma
}+\frac{g_{0}}{2}\sum\limits_{i}\sum \limits_{j\neq
i}f_{s}\frac{\rho _{ij}}{\rho _{0}}
 \\
&&+g_{\tau }\sum\limits_{i}\left( \frac{\rho
_{i}}{\rho _{0}}\right) ^{\eta }+\frac{C_{s}}{2}\sum\limits_{i}\sum%
\limits_{j\neq i}t_{i}t_{j}\frac{\rho _{ij}}{\rho _{0}}\left(
1-\kappa _{s}f_{s}\right),  \nonumber
\end{eqnarray}%
where
\begin{equation}
\rho_i=\sum\limits_{j\neq i} \rho_{ij} =\sum\limits_{j\neq i} \frac{1}{(4\pi \sigma _{r}^{2})^{3/2}}\exp [-\frac{(\mathbf{r}_{i}%
\mathbf{-r}_{j})^{2}}{4\sigma _{r}^{2}}],  \label{8}
\end{equation}%
\begin{equation}
f_{s}=\frac{3}{2\sigma _{r}^{2}}-\left( \frac{\mathbf{r}_{i}\mathbf{-r}_{j}}{%
2\sigma _{r}^{2}}\right) ^{2},  \label{9}
\end{equation}%
and $t_{i}=1$ for protons and $-1$ for neutrons respectively. The
parameters set IQ2 \cite{WN05} (see Table 1) is adopted in this
work. The Coulomb energy is written as the sum of the direct and
the exchange contribution, and the latter being taken into account
in the Slater approximation \cite{JCS51,JBK02}
\begin{equation}
U_{Coul}=\frac{e^{2}}{2}\int
\frac{\rho_{p}(\mathbf{r})\rho_{p}(\mathbf{r}^{\prime})}
{|\mathbf{r-r}^{\prime }|}d\mathbf{r}d\mathbf{r}^{\prime}
-e^{2}\frac{3}{4}\left( \frac{3}{\pi }\right) ^{1/3}\int \rho _{p}^{4/3}d%
\mathbf{r}.  \label{10}
\end{equation}%
%\begin{center}
%\text{Table 1. The Parameter set IQ2.}
%\end{center}
\begin{table}
 \caption{  Parameters set IQ2.}
\begin{tabular}{cccccccccc}
\hline Parameter & $\alpha $(MeV) & $\beta $(MeV) & $\gamma $ &$%
g_{0}$(MeVfm$^{2}$) & $ g_{\tau }$(MeV) & $\eta $ & $C_{s}$(MeV) &
$\kappa _{s}$(fm$^{2}$) & $\rho _{0}$(fm$^{-3}$) \\ \hline
IQ2 & -356 & 303 & 7/6 & 7.0 & 12.5 & 2/3 & 32.0 & 0.08 & 0.165 \\
\hline
\end{tabular}
\end{table}

To describe the fermionic nature of the N-body system and to
improve the stability of an individual nucleus, the phase space
occupation constraint method \cite{pp01} and the
system-size-dependent wave-packet width
$\sigma_{r}=0.09A^{1/3}+0.88$ fm \cite{WN05} are adopted. The
phase space occupation constraint is an effective approach to
improve the momentum distribution of nuclear system
\cite{WN02,pp01}. In this approach, the phase space occupation
number of each particle is checked at each time step. If the phase
space occupation number is larger than 1 for particle i, i.e.
$\bar f_{i}>1$, the momentum of the particle i are randomly
changed by a series of two-body elastic scattering between i and
its partner which guarantee that the total momentum and total
kinetic energy are conserved in the procedures. In the ImQMD
model, the new sample for the momenta of the particles is
constrained by the Pauli-blocking probability \cite{pp01} as in
the usual two-body collision process. Actually, the momenta of two
particles obtained in this way not only influence the motion of
particles of the system in this step but also the further more
steps. It is unknown whether the system will be in the most
suitable motion path. In this work, we perform one more step
further, i.e. we calculate the total energy of the system at step
$t$ and also the total energy $E(t+\Delta t)$ at the next time
step ($t+ \Delta t$) simultaneously. If the value of $E(t+\Delta
t)$ obviously deviates from that of $E(t)$, the two-body elastic
scattering procedure is re-executed. The number of times of the
re-executing process is small (about $0\sim4$) at each time step
for fusion reactions. This additional constraint can further
improve the stability of an individual nucleus (reducing the
spurious emission of nucleons), and is helpful for the study of
the formation process of the compound nuclei which lasts several
thousand fm/c or longer. We have checked that the total energy of
system is well conserved for thousands of fm/c with this new
procedure.

\begin{center}
\textbf{III. RESULTS}
\end{center}

In this section we first study the kinetic energies of a series of
nuclei. Then, we calculate the nucleus-nucleus potential in fusion
reactions based on the extended Thomas-Fermi approximation.

\begin{center}
\textbf{A. Kinetic energies of nuclei}
\end{center}

We first study the kinetic energy of a series of nuclei in the
ground state from $^{16}$O to $^{259}$No. Based on the extended
Thomas-Fermi (ETF)  approximation \cite{MBC85}, the kinetic energy
of a free Fermi gas can be expressed as
\begin{equation}  \label{11}
E_{k}=\frac{\hbar^{2}}{2m}\int\tau(\mathbf{r})d\mathbf{r}=c_k
\langle\rho\rangle^{2/3}+ \frac{\hbar^{2}}{2m}\frac{1}{36}\int
\frac {(\nabla \rho)^2}{\rho}d\mathbf{r}+ \dots
\end{equation}
with the kinetic energy density $\tau (\textbf{r})$ and the
coefficient
$c_k=\frac{\hbar^{2}}{2m}\frac{3}{5}(\frac{3\pi^2}{2})^{2/3} $.
With the help of the ETF form of the kinetic energy for Fermi gas
system in Eq.(12), we express the kinetic energy of an individual
nucleus in the ImQMD model as
\begin{equation}  \label{13}
E_{k}^{\rm ETF}\simeq c_{0}\sum\limits_{i}
 \rho_{i} ^{2/3}+  \frac{c_1}{\sum  \rho_{i}} \sum\limits_{i,j\neq i} f_{s}
\rho_{ij}  +c_{2}N.
\end{equation}
with $c_{0}=41.2$ MeVfm$^2$, $c_{1}=4.8$ MeVfm$^2$ and
$c_{2}=-1.0$ MeV  for IQ2, which are determined by fitting the
obtained kinetic energies of a series of nuclei with Eq.(4), see
Fig.2. $N$ is the particle number of the system under
consideration. The expressions of $\rho_{i}$, $\rho_{ij}$ and
$f_s$ are previously given in Eq.(9) and (10), respectively.  The
$c_{0}$ term of Eq.(13) represents the result of the Thomas-Fermi
approximation (see the $ \langle\rho\rangle^{2/3}
 $ term of Eq.(12)). The other terms give the corrections from the
finite system effect.

\begin{figure}
\includegraphics[angle=-0,width=1.0\textwidth]{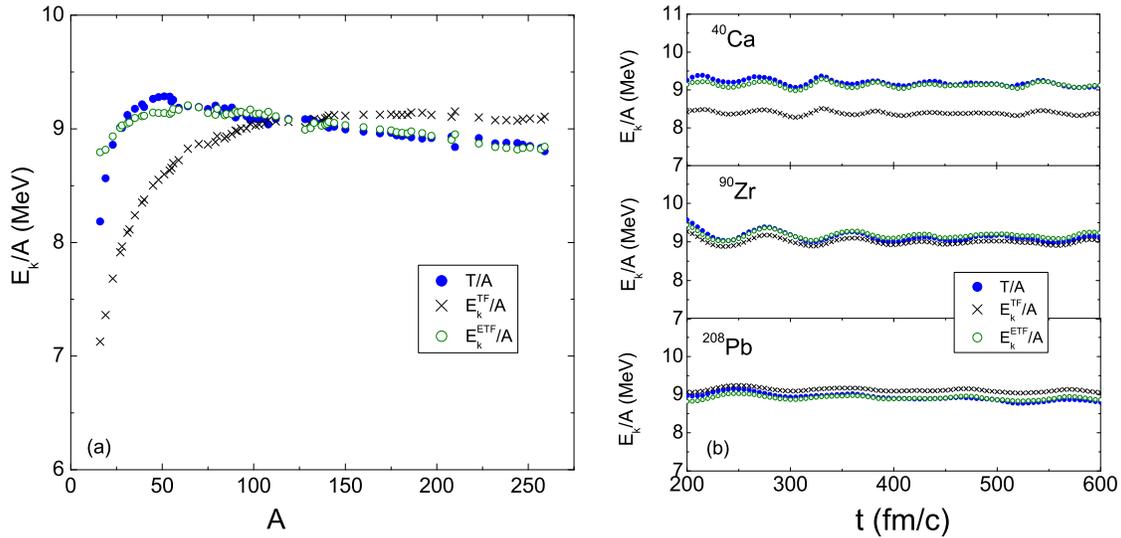}
 \caption{(Color online) (a) Time average of the kinetic energy per particle for a series of nuclei.
 The solid circle, the open circle and the crosses denote
the results with Eq.(4), (13) and (14), respectively. (b) The time
evolution of the kinetic energy per particle for $^{40}$Ca,
$^{90}$Zr, and $^{208}$Pb. }
\end{figure}

We show the time average of the kinetic energy per particle for a
series of nuclei in Fig. 2(a). Here we take 100 events for each
nucleus. The solid and open circles denote the results with Eq.(4)
and Eq.(13) which is based on the "ETF" approximation,
respectively. Here, "ETF" means that the form of Eq.(13) is
roughly obtained according to the extended Thomas-Fermi
approximation. The crosses denote the results with the "TF"
approximation
\begin{equation}
E_{k}^{\rm TF} \simeq c_{0}\sum\limits_{i} \rho_{i}^{2/3},
\end{equation}
in which the correction terms from the finite system effect are
not taken into account. From the figure one can see that for the
light nuclei ($A<50$), the calculated kinetic energies with the
"TF" approximation are much smaller than the values with Eq.(4),
while for heavy nuclei ($A>150$), the results with the "TF"
approximation are slightly larger than those with Eq.(4). Only for
the intermediate nuclei the results are in agreement with each
other. The calculated kinetic energy with the "ETF" approximation
is in good agreement with the values with Eq.(4) except for very
light nuclei. From Fig.2, one can see that the "ETF" approximation
can reasonably well describe the kinetic energy for finite nuclei.
The time evolution of the kinetic energies per particle for nuclei
$^{40}$Ca, $^{90}$Zr, and $^{208}$Pb are shown in Fig. 2(b). The
kinetic energies of these nuclei can be well described by the
Eq.(13) based on the "ETF" approximation.

\begin{center}
\textbf{B. Nucleus-nucleus potential in fusion reactions}
\end{center}

By using the ImQMD model, we can calculate the static and
dynamical Coulomb barriers. In the calculation of the static
Coulomb barrier which is based on the frozen density
approximation, the initial density distribution of the projectile
and target is adopted. In the calculation of the dynamical Coulomb
barrier, the realistic density distribution of the system which
changes with time due to the interaction between nucleons is used.
In this work, we concentrate on the calculation of the dynamical
fusion barrier. We study the dynamical nucleus-nucleus potential
$V$ based on the "ETF" approximation for the  kinetic energy.
According to the energy conservation, we have
\begin{equation}\label{16}
   E_{c.m.}=T_R+V+E^*+T_{oth},
\end{equation}
where $E_{c.m.}$ is the incident center-of-mass energy, $T_{R}$ is
the relative motion kinetic energy of two colliding nuclei, which
can be easily obtained in the ImQMD model since the position and
momentum of each nucleon can be followed at every time step in
this model, $E^*$ is the excitation energy, $T_{oth}$ is other
collective kinetic energy, such as neck vibration. When the
projectile and target nucleus are well separated ($R\gg R_1+R_2$),
the $E^*$ and $T_{oth}$ could be negligible which have been
checked by the time-dependent Hartree-Fock (TDHF) calculations
\cite{SEK1977, Umar06}, the nucleus-nucleus potential is thus
expressed as
\begin{equation}
V_1=E_{c.m.}-T_R.
\end{equation}
Where, $R_{1}$ and $R_{2}$ are the charge radii of the projectile
and the target nucleus, respectively, which are described by an
empirical formula $R_i=1.25A^{1/3}(1-0.2\frac{N-Z}{A})$ proposed
in \cite{BNP93}. After the di-nuclear system is formed
($R<R_{1}+R_{2}$), the nucleus-nucleus potential may be described
by a way like the entrance channel potential \cite{VYD02}
\begin{equation}
 V_2=E_{tot}(R)-\bar E_{1}-\bar E_{2},
\end{equation}
where $E_{tot}(R)$ is the energy of the composite system which is
strongly dependent on the dynamical density distribution of the
system obtained with the ImQMD model, $\bar E_{1}$ and $\bar
E_{2}$ are the time average of the energies of the projectile and
target nuclei, respectively. Here, the values of $\bar E_{1}$ and
$\bar E_{2}$ are obtained from the energies of the projectile
(like) and target (like) nuclei in the region $R_{T}<R<R_{T}+8$.
$R_T=R_{1}+R_{2}$ is the touching point. $R$ is the relative
distance between the two nuclei, which is a function of time.  In
the calculation of $E_{tot}(R)$, $\bar E_{1}$ and $\bar E_{2}$,
Eq.(13) that is a function of local density is used for the
description of the intrinsic kinetic energy of the system under
consideration.

In this work, we write the nucleus-nucleus potential as a smooth
function between $V_1$ and $V_2$,
\begin{equation}
V_{b}(R)=\frac{1}{2}{\rm erfc}(s)V_{2}+ [1-\frac{1}{2}{\rm
erfc}(s)]V_{1}
\end{equation}
where ${\rm erfc}(s)$ is the complementary error function
%\begin{equation}
%{\rm erfc}(s)=\frac{2}{\sqrt{\pi}}\int_s^{\infty} e^{-x^2} dx,
%\end{equation}
and
\begin{equation}
s=\frac{R-R_{T}+\delta}{\Delta R}
\end{equation}
with $\delta$=1fm, $\Delta R$=2fm. The obtained nucleus-nucleus
potential in Eq.(18) approaches to $V_1$ with the increase of the
separation distance between two nuclei. On the contrary,
$V_{b}(R)$ approaches to $V_2$ with the formation of the
di-nuclear system and the decrease of the distance between two
nuclei.
\begin{figure}
\includegraphics[angle=-0,width=0.8\textwidth]{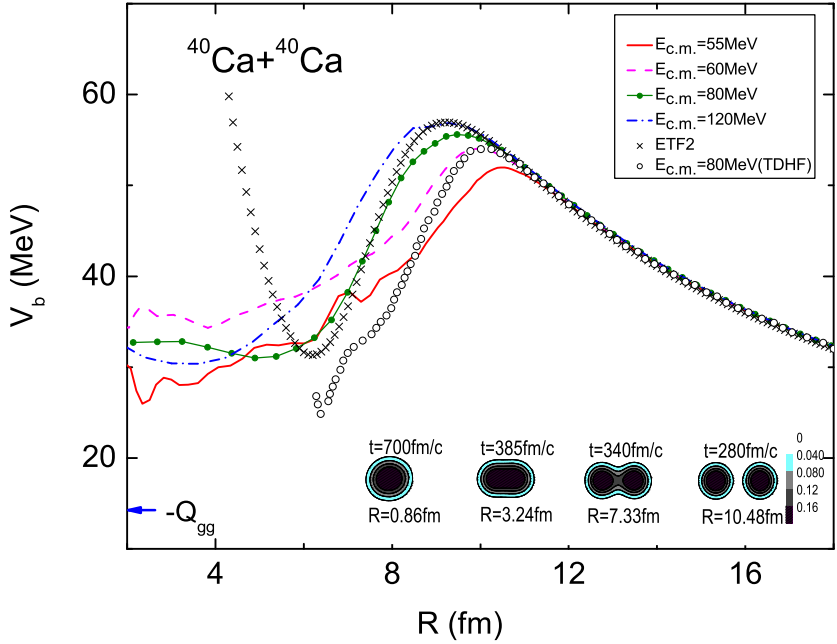}
 \caption{(Color online) The dynamical nucleus-nucleus potential of  $^{40}$Ca+$^{40}$Ca
 at different incident energies $E_{c.m.}$. The crosses denote
 the entrance channel potential with the Skyrme energy density funcitonal approach \cite{liumin06} which is based on frozen density approximation.
The open circle denotes the results of TDHF \cite{USA09} at
$E_{c.m.}$=80MeV. The inserted sub-figures denote the the density
distributions for this reaction at $E_{c.m.}$=80MeV and different
relative distances, respectively.}
\end{figure}
\begin{figure}
\includegraphics[angle=-0,width=0.6\textwidth]{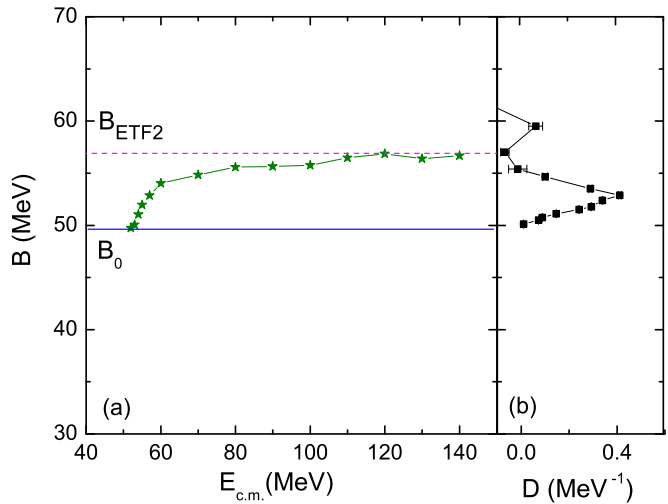}
 \caption{(Color online) (a) Barrier height $B$ for the reaction $^{40}$Ca+$^{40}$Ca
 at different incident energies  $E_{c.m.}$. The horizontal dashed
 and solid lines indicate the barrier height of the entrance channel potential based on the frozen density approximation \cite{liumin06}
and the lowest barrier $B_0$ extracted from the fusion excitation
function \cite{HAA84}, respectively. (b) The same as Fig.1(b). }
\end{figure}
To study the dynamical nucleus-nucleus potential, we create 500
reaction events for head-on collision of $^{40}$Ca+$^{40}$Ca at
several center-of-mass (c.m.) energies ranging from $E_{c.m.}=52$
to 140 MeV. For each event, we evolve the reaction system for a
time of 700 fm/c. The distance between the projectile and target
at the initial time is set to 30 fm for this reaction. The
scattering events at $t=700$ fm/c are not involved in the
calculation of the nucleus-nucleus potential. Fig. 3 shows the
obtained dynamical nucleus-nucleus potential at different incident
energies $E_{c.m.}$. The corresponding density distributions at
$E_{c.m.}$=80MeV and different relative distances are also shown
in the sub-figures. Fig. 4  shows the average fusion barrier
height $B$ for the reaction $^{40}$Ca+$^{40}$Ca at different
$E_{c.m.}$. From the Fig. 3  and Fig. 4  we can see that the
dynamical barriers depend on the incident energies. At energies
around the Coulomb barrier, the dynamical barrier increases
rapidly with the incident energy. With the further increase of the
incident energy, the dynamical barrier approaches to the barrier
height of the entrance channel potential (56.9 MeV) which is
obtained with the Skyrme energy density functional together with
the frozen density approximation \cite{liumin06}. This trend has
also been found in \cite{TJL09,KW08}. When the incident energy
decreases gradually and down to $E_{c.m.}=55$ MeV, the height of
dynamical barrier falls to 52.0 MeV. With lower incident energy,
the height of the dynamical barrier approaches to about 50 MeV
which is close to the height of the lowest barrier $B_0$ extracted
from the fusion excitation function. In addition, it is
encouraging that the obtained barrier height and the depth of the
fusion pocket in this work are comparable with the results of the
TDHF calculations \cite{USA09}. The depth of the fusion pocket is
about 25 MeV for this reaction system.

At very short distances between two nuclei, it is thought that the
$Q$ value of the fusion system may provide information of the
nucleus-nucleus potential. One commonly defines the excitation
energy for a reaction from the expression
\begin{equation}\label{21}
-Q_{gg}=E_{c.m.}-E^*,
\end{equation}
where $Q_{gg}$ is the mass difference between the two initial
nuclei and the combined system in its ground state. From Eq.(15)
and Eq.(17), one can find that when the compound nucleus is well
formed and the collective motion can be negligible one gets
$V(R=0)=-Q_{gg}$, which is the result of the adiabatic
nucleus-nucleus potential \cite{Zag}. For the reaction
$^{40}$Ca+$^{40}$Ca, we have $-Q_{gg}$=14.3 MeV. Actually, because
the expression (20) is correct relative to the ground state of the
composite system, it does not accurately describe the excitation
energy relative to other intermediate transition states formed
during the collision \cite{USA09}, the nucleus-nucleus potential
obtained from the ImQMD model and the TDHF \cite{USA09} at short
distance does not exactly reach the value $-Q_{gg}$ since the
composite system formed during the collision is far from a ground
state.

\begin{figure}
\includegraphics[angle=-0,width=0.9\textwidth]{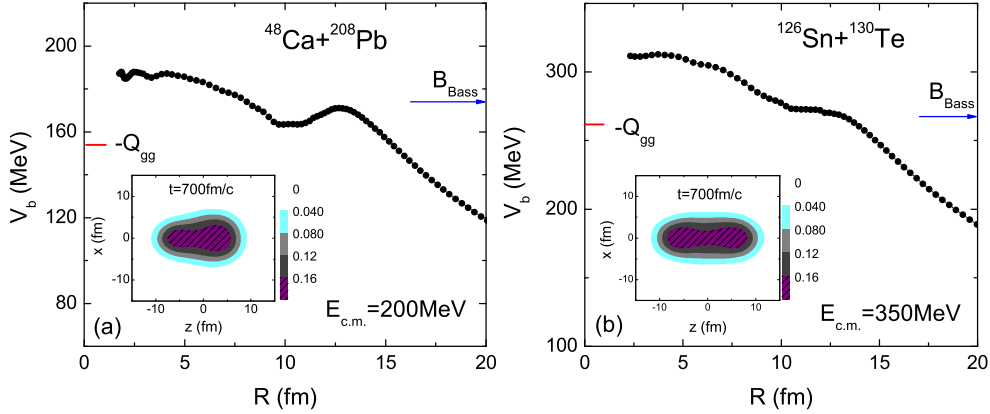}
 \caption{(Color online) Dynamical nucleus-nucleus potentials for
 the reactions $^{48}$Ca+$^{208}$Pb and $^{126}$Sn+$^{130}$Te at
 an incident energy $E_{c.m.}$=200MeV and $E_{c.m.}$=350MeV,
 respectively. The initial distance is set to 40 fm.
 The arrows denote the corresponding Bass barriers.}
\end{figure}

With the same approach we studied the nucleus-nucleus potential
for the reactions $^{48}$Ca+$^{208}$Pb and $^{126}$Sn+$^{130}$Te
at energies above the Coulomb barrier. The corresponding values of
$-Q_{gg}$ for these two reactions are 153.8 and 261.2 MeV,
respectively. These two reactions lead to the same compound
nucleus $^{256}$No. Fig. 5 shows the calculated nucleus-nucleus
potentials for these two reactions. The arrows denote the Bass
barriers. From Fig. 5 we can see that the obtained barrier heights
are close to the corresponding Bass barriers. The depth of the
fusion pocket (about 7 MeV) for $^{48}$Ca+$^{208}$Pb becomes much
shallower than that of $^{40}$Ca+$^{40}$Ca  (about 25 MeV) and the
fusion pocket for $^{126}$Sn+$^{130}$Te almost disappears, which
indicates that quasi-fission could easily occur in heavy fusion
process especially for the more symmetric systems. Furthermore, we
find that the nucleus-nucleus potentials for the reactions
$^{48}$Ca+$^{208}$Pb and $^{126}$Sn+$^{130}$Te at short distances
are much higher than the value of $-Q_{gg}$ and even higher than
the Coulomb barrier, which is quite different from the case of
$^{40}$Ca+$^{40}$Ca. These calculations indicate: 1) additional
incident energy (so-called extra-push energy \cite{Shen}) beyond
the energy to overcome the Coulomb barrier may be required to form
the compound nucleus for heavy fusion system and 2) the process of
nucleon transfer between the projectile (like) and the target
(like) could last for a period of time due to the appearance of
the fusion pocket in di-nuclear system, which is the basic
assumption of the di-nuclear system (DNS) model \cite{Ada98}. To
see the fusion path, we also show the corresponding density
distributions of the composite systems at $t=700 fm$ in the
inserted sub-figures. One can see that the corresponding compound
nuclei are not well formed at $t=700 fm$ for these two heavy
fusion systems. The strongly deformed composite systems or called
di-nuclear systems are formed at about $t=350$ fm and can last
hundreds even thousands fm/c for heavy fusion system, which is
quite different from the case of light system such as
$^{40}$Ca+$^{40}$Ca in which the spherical composite system is
well formed at $t=700 fm$ with the incident energies above the
Coulomb barrier (see Fig.3). For $^{126}$Sn+$^{130}$Te, the
composite system tends to undergo quasi-fission or fission. In
Fig.6 we show the capture cross sections of these two reactions.
The solid and open circles in Fig.6(a) denote the experimental
data of $^{48}$Ca+$^{208}$Pb in Ref.\cite{Prok03} and
\cite{Pach92}, respectively. The solid curves denote the results
of an empirical barrier distribution approach which is based on
the Skyrme energy-density functional together with the extended
Thomas-Fermi (ETF) approximation \cite{liumin06,Wang06}. The solid
squares denote the results of the ImQMD model with IQ2 and the
error bars denote the corresponding statistical errors. For the
reaction $^{48}$Ca+$^{208}$Pb, the experimental data at energies
above the Coulomb barrier can be reproduced acceptably well by the
ImQMD model. Because the ImQMD model has difficulties to deal with
the shell effects, the capture cross sections of
$^{48}$Ca+$^{208}$Pb at sub-barrier energies  can not be described
well. For $^{126}$Sn+$^{130}$Te, the calculated capture cross
sections with the ImQMD model are comparable with the results of
the empirical barrier distribution approach.

\begin{figure}
\includegraphics[angle=-0,width=0.9\textwidth]{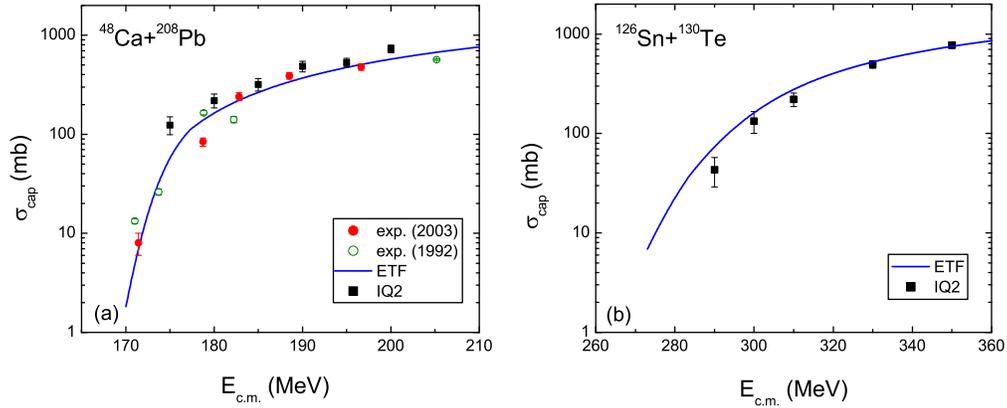}
 \caption{(Color online) Capture cross sections for
 the reactions $^{48}$Ca+$^{208}$Pb and $^{126}$Sn+$^{130}$Te.
 The open and solid circles in (a) denote the experimental
data of $^{48}$Ca+$^{208}$Pb \cite{Prok03,Pach92}. The solid
curves denote the results of an empirical barrier distribution
approach proposed in \cite{liumin06,Wang06}. The solid squares
denote the results of the ImQMD model and the error bars denote
the corresponding statistical error.}
\end{figure}

\begin{center}
\textbf{IV. SUMMARY}
\end{center}

In summary, the kinetic energies of a series of nuclei have been
studied with the ImQMD model together with the extended
Thomas-Fermi (ETF) approximation which gives accurate results for
finite nuclear system, especially for the light and heavy nuclei.
Furthermore, with the "ETF" approximation for the kinetic energies
we have studied the dynamical Coulomb barrier of the reaction
$^{40}$Ca+$^{40}$Ca  at different incident energies. The results
show that the dynamical Coulomb barrier strongly depends on the
incident energy. With the increase of the incident energy, the
dynamical Coulomb barrier increases gradually and approaches to
the entrance channel potential which is based on the frozen
density approximation. The height of dynamical Coulomb barrier
decreases with the decrease of the incident energy and approaches
to the lowest barrier extracted from the fusion excitation
function. The behavior of nucleus-nucleus potential at short
distances for heavy system is obviously different from that of
light systems. For heavy fusion systems, the depth of the fusion
pocket becomes much shallower and the nucleus-nucleus potential at
short distances are higher than the adiabatic potential. The
capture cross sections for $^{48}$Ca+$^{208}$Pb and
$^{126}$Sn+$^{130}$Te have also been studied with the ImQMD model.
The calculated results are comparable with the results of the
empirical barrier distribution approach. A systematic study of
heavy fusion systems, such as the calculation of the potential
energy surface of the composite system in the fusion process as a
function of mass-asymmetry and the distance between two nuclei is
in progress with the shell effects being taken into account.

\begin{center}
\textbf{ACKNOWLEDGEMENTS}
\end{center}
This work is supported by the National Natural Science Foundation
of China, Nos 10875031, 10847004 and 10979024, and the Innovation
Project of Guangxi Graduate Education, No. 2009106020702M36.

\end{document}